\documentclass{article}

\usepackage{PRIMEarxiv}
\usepackage{amsmath}
\usepackage[utf8]{inputenc} 
\usepackage[T1]{fontenc}    
\usepackage{hyperref}       
\usepackage{url}            
\usepackage{booktabs}       
\usepackage{amsfonts}       
\usepackage{nicefrac}       
\usepackage{microtype}      
\usepackage{lipsum}
\usepackage{fancyhdr}       
\usepackage{graphicx}       
\graphicspath{{media/}}     

\title{The Information Theory of Self-Organization Phenomena in Thermal Systems}

\author{
  Hongzheng Liu \\
  Independent \\
  China \\
  \texttt{\{weiyouyeyu\}@foxmail.com} \\
   \And
  Zhiyue Wu \\
  Independent \\
  China \\
  \texttt{\{Zhiyuewu70\}@gmail.com} \\
}

\begin{document}
\maketitle

\begin{abstract}
This paper revisits Brownian motion from the perspective of Information Theory, aiming to explore the connections between Information Theory, Thermodynamics, and Complex Science. First, we propose a single-particle discrete Brownian motion model (SPBM). Within the framework of the maximum entropy principle and Bayesian inference, we demonstrate the equivalence of prior information and constraint conditions, revealing the relationship between local randomness and global probability distribution. By analyzing particle motion, we find that local constraints and randomness can lead to global probability distributions, thereby reflecting the interplay between local and global dynamics in the process of information transfer. Next, we extend our research to multi-particle systems, introducing the concepts of "Energy as Encoding" and "Information Temperature" to clarify how energy distribution determines information structure. We explore how energy, as not only a fundamental physical quantity in physical systems but also an inherently informational one, directly dictates the prior probability distribution of system states, thus serving as a form of information encoding. Based on this, we introduce the concept of "Equilibrium Flow" to explain the self-organizing behavior of systems under energy constraints and Negative Information Temperature. By proving three theorems regarding Equilibrium Flow systems, we reveal the criticality of Self-Organization, energy-information conversion efficiency, and the characteristic that event occurrence probabilities follow the Fermi-Dirac distribution. Through theoretical analysis and theorem proofs, we offer new perspectives for understanding the dynamics of Complex Systems, enriching the theoretical framework of Information Theory, Thermodynamics, and Complex Science, and providing a new theoretical basis for further research in related fields.
\end{abstract}


\section{Introduction}
Brownian motion, as a fundamental random process, plays a significant role in the fields of statistical Physics, Financial Mathematics, and Biophysics \cite{uhlenbeck1930theory, kubo1966fluctuation, van1992stochastic, del2016control}. It provides a theoretical basis for understanding the random motion of particles, diffusion phenomena, and the connections between microscopic and macroscopic behaviors, contributing significantly to the description of the dynamics of Complex Systems \cite{xu2020solving, gorenflo2020mittag, lacasce2008statistics}. Information theory offers a powerful theoretical framework for communication and computation by quantifying the processes of information transmission and processing \cite{bennis2018ultrareliable, jack2015human, o2016learning}. By introducing concepts such as entropy and mutual information, Information Theory plays a key role in understanding system uncertainties, data compression, and signal processing, profoundly impacting various disciplines \cite{li2017feature, jack2017toward, gopalakrishnan2018gut, vuong2024better}.
Although Thermodynamics, Information Theory, Brownian motion and Stochastic Physical Process share commonalities in handling energy, information, and uncertainties within systems, their deeper connections remain underexplored \cite{bennett1982Thermodynamics, deffner2016foundations, dykman2012fluctuating, berthier2011theoretical}. Specifically, a unified description of energy distribution, interactions, and system constraints in physical systems from the perspective of Information Theory faces numerous challenges \cite{hanggi2009artificial, borzuei2020role, aldosari2021predicting}. 

To address this, our work aims to revisit Brownian motion from the perspective of Information Theory, establishing a theoretical framework from single-particle to multi-particle systems to unify the exploration of the connections between Information Theory and Thermodynamics.
In our work, we first focus on a simple single-particle discrete Brownian motion model (SPBM), discretizing one-dimensional space into an infinite number of equidistant points where the particle can move, forming a discrete random walk. Based on this model, we prove the equivalence of prior information and system constraints within the framework of the maximum entropy principle and Bayesian inference. We introduce a Bayesian mechanism that integrates prior information and probabilistic updates, revealing how constraints influence information transfer within the system and elucidating the interplay between local dynamics and global behaviors.
Next, we extend our research to multi-particle systems and propose the concept of "Energy as Encoding." We emphasize that energy, beyond being a fundamental physical quantity in physical systems, inherently possesses informational properties, directly determining the prior probability distribution of system states and serving as a form of information encoding. By linking Boltzmann distribution and information entropy, we clarify how energy distribution shapes information structures, unveiling the deeper connections between energy and information. Moreover, we introduce the concept of "Information Temperature," interpreting temperature as a measure of the frequency and intensity of information exchange within a system. We unify entropy, energy, and temperature under one framework, highlighting their interrelationships.
Building on this, we propose the concept of "Equilibrium Flow" to explain self-organizing behavior in open systems. We analyze how systems transition from disorder to order through energy input and entropy output under conditions of Negative Information Temperature, showcasing the interplay of energy, information, and entropy in Complex Systems. Finally, we apply our framework to the training process of large language models (LLMs). From the perspective of information principles, we illustrate the self-organizing behaviors observed during LLM training, explaining how models achieve entropy reduction or maintain stability through continuous energy input and negentropy under energy constraints, forming a dynamic equilibrium. By proving theorems regarding Equilibrium Flow systems, we further propose that the probability of events occurring under such systems follows the Fermi-Dirac distribution. The main contributions of this study are as follows:
    \begin{enumerate}  
        \item We propose SPBM and demonstrate the equivalence of prior information and system constraints within the framework of the  Bayesian inference, revealing the interplay between local and global dynamics. 
        \item We introduce the concepts of "Energy as Encoding" and "Information Temperature," clarifying how energy distribution shapes system information structures and exploring the profound connections between Information Theory and Thermodynamics. 
        \item We propose the concept of "Equilibrium Flow" to explain self-organizing behavior in open systems, analyzing the dynamic characteristics of systems under conditions of Negative Information Temperature and applying this to explaining the training process of LLMs. 
        \item Through theoretical analysis and theorem proofs, the characteristics of event occurrence probabilities following the Fermi-Dirac distribution were proposed, along with the energy-information conversion efficiency in Equilibrium Flow systems. Provide new perspectives for understanding the dynamics of Complex Systems and unveiling the interactions of energy, information, and entropy within these systems. 
    \end{enumerate}
Section \ref{SPBM_section} introduces the definition and basic properties of the SPBM, discussing the equivalence of prior information and constraints as well as the interplay between local and global dynamics. Section \ref{Multi_Brownian} extends the research to multi-particle systems, proposing the concepts of "Energy as Encoding" and "Information Temperature" and analyzing the relationship between energy distribution and information structure.Furthermore, we propose concept of "Equilibrium Flow," uncovering self-organizing behaviors in open systems and further proposing that event probabilities under Equilibrium Flow systems follow the Fermi-Dirac distribution, illustrated by examples such as the training of LLMs. 

\section{Related work}

\subsection{Classical Theory of Brownian Motion}
The classical theory of Brownian motion, established in the early 20th century, has been fundamental in the development of statistical Physics and Stochastic Processes. Robert Brown first observed the random motion of particles, while Albert Einstein and Marian Smoluchowski independently provided theoretical models in 1905, explaining this phenomenon through molecular collisions\cite{einstein1905movement}. This theoretical framework led to the development of diffusion equations and the Wiener process, formalizing Brownian motion as a continuous-time stochastic process \cite{schilling2014brownian, ikeda2014stochastic}. Extensions and refinements of this theory have enabled modeling in various fields, including Physics, Chemistry, and Finance \cite{coelho2019resolving, grzybowski2009chemistry, kou2007jump}. Despite the robustness of these models, classical studies have predominantly focused on particle trajectories and probabilistic distributions without deeply exploring the underlying informational aspects of such stochastic behaviors \cite{simonton2003scientific}. Recent advancements in stochastic Thermodynamics \cite{seifert2008stochastic, parrondo2015Thermodynamics} and the application of information-theoretic principles \cite{ito2018stochastic, barato2014unifying} have motivated a reexamination of Brownian motion to address its informational and energetic properties in Complex Systems \cite{seifert2012stochastic}. These emerging perspectives highlight the need to unify traditional physical models with modern concepts of information and energy constraints \cite{dyson1962statistical, sasaki2014quantum}.

\subsection{Integration of Information Theory and Statistical Mechanics}
Information theory, pioneered by Claude Shannon in 1948, revolutionized the understanding of communication and data processing \cite{shannon2001mathematical}. Since its inception, researchers have sought to apply information-theoretic approaches to physical and statistical systems to analyze uncertainty and entropy \cite{thurner2017three, bekenstein2004black}. The integration of Information Theory with statistical mechanics has proven particularly fruitful, enabling a deeper understanding of the entropy-energy relationship and thermodynamic irreversibility 
\cite{jaynes1957information, gray2011entropy}. Jaynes' work on the maximum entropy principle \cite{jaynes1957information_2} established a pivotal link, suggesting that probability distributions in thermodynamic systems can be derived by maximizing entropy subject to known constraints \cite{shore1980axiomatic}. This principle has been further expanded to describe complex phenomena in non-equilibrium systems and information transmission \cite{dyke2010maximum, dewar2013beyond}. Recent research has focused on Bayesian approaches, demonstrating how prior knowledge and constraints influence system behavior and probability updates \cite{paulson2019bayesian, he2024large}. These studies illustrate how entropy serves as both a measure of information content and a driver of thermodynamic processes, providing a framework for understanding the intersection of physical dynamics and information transmission \cite{von2011bayesian, gray2020bayesian}.

\subsection{Relationship Between Energy and Information}
Energy plays a critical role in determining the dynamics and state evolution of physical systems, yet it also holds profound implications from an informational standpoint \cite{tribus1971energy, ortega2013Thermodynamics}. The notion that energy distribution carries and encodes information has been explored in contexts ranging from thermodynamic entropy to quantum information processing \cite{poplavskiui1975thermodynamic, wilde2018energy}. Brillouin's concept of "negentropy" emphasized the informational content associated with organized states \cite{brillouin1953negentropy}. Further, energy-based models such as Boltzmann machines have illustrated how energy landscapes encode probabilistic distributions, forming a basis for generative models and optimization algorithms \cite{zhang2018overview, hinton2007boltzmann}. By framing energy as a carrier of information, researchers have established connections between the physical state of a system and its informational structure, exemplifying how thermodynamic states can be inferred through probabilistic and entropic measures \cite{machta1999entropy}, Thermodynamic States and Information Theory: A Probabilistic Approach. This approach has been extended to understand complex, multi-particle systems and their emergent properties \cite{machado2013complex}. We introduction of concepts such as "Informational Temperature" seeks to quantify the frequency and intensity of information exchange, further elucidating the interplay between energy and informational dynamics. 

\subsection{Self-Organization and Non-Equilibrium Thermodynamics}
Self-organization is a phenomenon observed in many Complex Systems, where global order arises spontaneously from local interactions \cite{banzhaf2009self}. Non-equilibrium Thermodynamics provides a theoretical basis for describing such behavior, focusing on how systems evolve through exchanges of energy and entropy with their environments \cite{kleidon2004non}. Prigogine's work on dissipative structures demonstrated how systems far from equilibrium can develop stable patterns through the continuous exchange of energy \cite{prigogine2018order, prigogine1985self}. This dynamic balance of energy input and entropy output is closely tied to Information Theory, as it captures the exchange and reduction of uncertainty within the system \cite{haken2016information, zu2020information}. Recent studies have examined the role of "Negative Temperatures," characterizing states where higher-energy configurations are more probable \cite{swendsen2016negative}. Further advancements have highlighted how self-organizing systems can achieve entropy reduction through energy-driven processes \cite{guerin2004emergence}. This framework aligns with recent work in Complex Systems, offering a robust understanding of the coupling between energy, entropy, and information flow \cite{holland1992complex, zurek2018complexity, baudin2023observation}. Further, based on the physical characteristics of Negative Temperature, specifically the reversal of high-energy states leading to frequent low-probability events in phenomena, we introduce the concept of "Negative Information Temperature." This concept explains the occurrence of high-energy configurations in certain systems and extends the traditional understanding of temperature in Thermodynamics and Information Theory, providing a novel framework for analyzing Complex Systems exhibiting non-equilibrium behavior.

\section{Single-Particle Discrete Brownian System}
\label{SPBM_section}
Whether in Complex Systems or simple phenomena, their core essence is deeply related by random processes. From the perspective of macroscopic random processes, they are fundamentally statistical aggregates of countless microscopic thermal motions interacting with one another. We refer to such phenomena as "physical truncation," which describes the transition from the infinite degrees of freedom of microscopic systems to the finite manifestations of macroscopic systems. In this process, numerous microscopic thermal molecules interact in complex and random ways to form a certain statistical behavior. While the details of this behavior are difficult to directly observe on a macroscopic scale, its randomness manifests through the truncation process in specific patterns.It is not merely a loss of information but also a reorganization and macroscopic expression of the randomness inherent in the system. This perspective reveals a profound truth: randomness pervades both simple and Complex Systems. The randomness of microscopic thermal molecular interactions is truncated to present orderly random behavior patterns at the macroscopic scale. For instance, the essence of Brownian motion lies in the statistical result of macroscopic particles or idealized particles being randomly struck by countless thermal molecules. This randomness manifests as irregular motion on a macroscopic scale. Therefore, to gain deeper insights into the nature of Complex Systems, we choose to study Brownian motion as a simple model, analyzing its regularities to uncover the universal principles of random processes in Complex Systems.

\subsection{Defination}
To thoroughly explore the interaction between constraints and randomness, we focus on one-dimensional space and SPBM as our research object. This simplification allows us to concentrate on key analytical issues. Specifically, we discretize one-dimensional space into an infinite number of equidistant points, with each point representing a degree of freedom, and particles can only move between these discrete points, forming the SPBM model. In this model, the states and movement rules of particles are set as follows:
\begin{enumerate} 
    \item \textbf{State of system degrees of freedom}: Each position can only be in one of the following two states: 
    \begin{itemize} \item \emph{Steady state} (state 0): Indicates that the position is not occupied by a particle. 
    \item \emph{Excited state} (state 1): Indicates that the position is occupied by a particle. \end{itemize} 
    \item \textbf{Single-particle constraint}: There exists only one particle in the system at any given time, avoiding the complexity of multi-particle interactions and allowing us to focus on single-particle motion behavior. Mathematically, this constraint can be expressed as: 
\end{enumerate}


\begin{equation}
    \sum_{x} s(x, t) = 1, \quad \forall t
    \label{SingleParticleConstraint}
\end{equation}
Here,\( s(x, t) \in \{0, 1\} \)represents the state of position \( x \) at time \( t \), with\( s(x, t) = 1 \)indicating that a particle is at position $x$ at time $t$. This constraint ensures the conservation of total system energy, as the presence of a particle contributes to system energy, while the particle's movement does not change the total system energy.

\textbf{Adjacent displacement principle}: Particles can only move to adjacent positions within each time step, meaning that a particle can only move from position 
\( X(t) \) to \( X(t \pm \Delta x) \) where \( \Delta x \) is the spatial step length. This principle limits the maximum displacement of a particle per time step, ensuring the continuity and finiteness of physical motion. Mathematically, this principle is expressed as:
\begin{equation}
    \{\begin{aligned} &P(X_{t+\Delta t} = x \pm \Delta x \mid X_t = x) > 0 \ &P(X_{t+\Delta t} \notin { x \pm \Delta x } \mid X_t = x) = 0 \end{aligned}. 
    \label{ConstraintConditions}
\end{equation}

\textbf{Randomness of motion}: To ensure the model's unbiasedness and symmetry, we assume that the particle moves left or right by one spatial step with equal probability
\begin{equation}
    P(X_{t+\Delta t} = x + \Delta x \mid X_t = x) = P(X_{t+\Delta t} = x - \Delta x \mid X_t = x) = \dfrac{1}{2}
    \label{EqualProbability}
\end{equation}
Combining the above conditions, the motion of the particle can be represented as:

\begin{equation} 
X(t + \Delta t) = X(t) + \Delta X 
\end{equation} 

\begin{equation} 
\Delta X = \begin{cases} + \Delta x, & \text{with probability } \dfrac{1}{2}, \\ -\Delta x, & \text{with probability } \dfrac{1}{2}. \ \end{cases} 
\end{equation}
Therefore, the recurrence relation for the particle's probability distribution can be expressed as:

\begin{equation} P(x, t + \Delta t) = \dfrac{1}{2} P(x - \Delta x, t) + \dfrac{1}{2} P(x + \Delta x, t) \label{Recursion} \end{equation}

\subsection{Priors and Conditions Equivalence}
In the Bayesian statistical framework, prior information and constraint conditions both describe our understanding of the system prior to observing data. Prior information reflects our initial beliefs about the system's state, while constraint conditions impose limitations on possible system behaviors. In the context of SPBM, our aim is to demonstrate the equivalence between prior information and constraint conditions. This equivalence implies that specifying a prior probability distribution is equivalent to imposing certain constraints on the system's behavior.

Consider a random walk in one-dimensional discrete space, where space is discretized into a series of equidistant points. The constraint condition for the system is that within a time step\(\Delta t\)the particle can only move from its current position \(x\) to an adjacent position\(x\pm a\). The particle's possible positions are in the set \(\mathcal{X}=\{\cdots,x_{-2},x_{-1},x_0,x_1,x_2,\cdots\}\). According to \eqref{ConstraintConditions}, the prior condition within this framework is represented by the prior probability distribution$P(X_{t+\Delta t}|X_t)$, which reflects our understanding of the system state prior to observation. The constraint assigns positive probabilities to positions where the particle may move, while assigning zero probabilities to all other positions:

\begin{equation}
\left\{
\begin{aligned}
&P(X_{t+\Delta t}=x'|X_t = x)=\frac{1}{2},\text{ when }x' = x + a\text{ or }x' = x - a\\
&P(X_{t+\Delta t}=x'|X_t = x)=0,\text{ in other cases}\\
&\sum_{x'}P(X_{t+\Delta t}=x'|X_t = x)=1
\end{aligned}
\right.
\label{prior}
\end{equation}
Sufficiency: If condition \(A\) holds, then condition \(B\) must also hold, i.e., \(A\rightarrow B\). Necessity: If condition \(B\)  holds, then condition \(A\) must also hold, i.e., \(B\rightarrow A\) To prove the equivalence of prior conditions and constraint conditions in SPBM, we must show that prior and constraint conditions are sufficient and necessary for each other. In the case of sufficiency, we must demonstrate that if a state has non-zero probability in the prior distribution, i.e., \(P(X_{t+\Delta t}=x')>0\), then this state must satisfy the prior condition \(P(X_{t+\Delta t}=x'|X_t = x)=\frac{1}{2}\).
Suppose a state \(X_{t+\Delta t}=x'\) has a non-zero probability in the prior distribution, i.e., \(P(X_{t+\Delta t}=x'|X_t=x)>0\). According to the prior distribution definition\eqref{prior} follows that \(x'=x+a\) or \(x'=x - a\) satisfies the condition, indicating that \(x'\) adheres to the adjacent displacement principle, as the particle can only move from its current position to adjacent positions. Additionally, since only one particle occupies a position at any given time in the system, each time step in the prior distribution involves only one particle position. Thus, satisfying the prior condition necessarily satisfies the constraint condition.
For the constraint condition: If \(X_{t+\Delta t}=x'\) satisfies the constraint condition, then must hold. The prior distribution is then defined as \eqref{prior} Given that \(x'=x\pm a\) according to the definition of the prior distribution, \(P(x+\Delta t=x)>0\). Therefore, a state that satisfies the constraint condition has a non-zero probability in the prior distribution. Thus, in the SPBM model.

\begin{center}
    \normalsize{Theorem 1:}  \textbf{Prior information and Constraint conditions are Equivalent}
\end{center}


We have pointed out the equivalence of prior information and constraint conditions in the SPBM model. This equivalence not only simplifies the expression of the local dynamic behavior of the model, but also provides a unified theoretical framework for analyzing the system behavior from a global perspective. The introduction of the Bayesian mechanism further expands this perspective, and through the dynamic adjustment of prior information, an effective connection is established between local constraints and global distributions.

\subsection{Local and Global interaction}
In this section, we explore the interaction between local dynamics and global behavior in the SPBM model from an information perspective. We indicate that

\begin{center}
    \normalsize{\textbf{Theorem 2:}} \textit{The interaction between the global and local levels facilitates the evolution of stochastic processes.}
\end{center}

Namely, the change in information entropy during the system's stochastic evolution is jointly determined by the conditional entropy at the microscopic level and the mutual information at the macroscopic level.

\subsubsection{Local-Global Relationships from an Information Perspective.}
The Brownian motion of a particle is a fundamental random process that not only reflects the random motion properties of the particle but also reveals the dynamic transmission and redistribution of information in space. The introduction of the adjacent displacement principle restricts the range of particle motion, leading to an orderly and localized transmission of information, resulting in a non-uniform distribution of probabilities over space and time. At the local level, particle motion is constrained by physical laws and randomness, leading to local updates of the probability distribution. The entropy of the system at time $t$ and the joint entropy is $H(x,y)=H(x)+H(y)- I(x,y)$, where $\Delta H$ can be transformed into

\begin{equation}
    \Delta H = H_{cond} + I(X_t, X_{t+\Delta t}) - H(t).
    \label{Mutual Information trans}
\end{equation}
As Brownian motion proceeds over an infinite time, the expected value of Brownian motion equals zero, meaning the particle's expectation should be at the origin. However, in practice, observed particles undergoing Brownian motion often drift away from the origin and perform diffusion. By analyzing this relationship between local and global interactions through quantitative formulas \eqref{Mutual Information trans}, we find that physical laws not only constrain possible states of the system but also influence the transmission and update of information through prior distributions. This indicates that each particle movement step represents not only a physical location change but also a dynamic update of system information. Using the Bayesian mechanism, we quantify the process of information transfer and reorganization, understanding how information propagates within the system. Thus, Brownian motion maintains not only its characteristic random diffusion but also a structured propagation of information across space and time.
Specifically, each particle movement corresponds to a recalculation of information entropy and an adjustment of distribution, reflecting changes in system uncertainty and information flow. This dynamic updating process highlights the central role of information in random processes, showing that information not only affects the current state of the system but also its future evolution paths. Described through the Bayesian mechanism, Brownian motion can be seen as a "constrained random process," where randomness is guided by prior information, giving structure to the process. In a system with priors, global behavior exhibits randomness, while local behavior displays structure.
From a local information perspective, the change in information entropy is given by

\begin{equation*}
    \frac{dH(t)}{dt}=-\sum_{x}\frac{d(P(x,t)\ln P(x,t))}{dt}
\end{equation*}

According to the derivation rule of products and the sum of probabilities is always 1, and the derivative with respect to time is 0, which is

\begin{equation}
    \frac{dH(t)}{dt}=-\sum_{x}\frac{dP(x,t)}{dt}\ln P(x,t)-\sum_{x}\frac{dP(x,t)}{dt} = -\sum_{x}\frac{dP(x,t)}{dt}\ln P(x,t)
\end{equation}
Due to the constraints of single-particle Brownian motion, as time increases, the likelihood of particles appearing at different positions grows. In Brownian motion, the random changes in particle position cause probabilities at some positions to increase and at others to decrease. Let the set of positions with increasing probability be \(A=\{x_{i_1},x_{i_2},\cdots,x_{i_m}\}\), then for \(x_{i_j}\in A\), \(\frac{dP(x_{i_j},t)}{dt}>0\). Let the set of positions with decreasing probability be \(B=\{x_{k_1},x_{k_2},\cdots,x_{k_{n-m}}\}\), then for \(x_{k_l}\in B\), \(\frac{dP(x_{k_l},t)}{dt}<0\). Due to the randomness of Brownian motion, particle positions become more dispersed, meaning that although probabilities decrease at some positions, more positions will have non-zero probabilities, with new probabilities leading to an overall increase in entropy. Positive contributions arise from more positions and larger changes in probability, while negative contributions come from fewer positions and smaller changes in probability. Thus, on the whole, positive contributions exceed negative contributions, implying \(\frac{dH(t)}{dt}>0\), showing that entropy increases with time. 

\section{Multi-Particle Brownian System}

\label{Multi_Brownian}
In the previous chapters, we provided a detailed introduction to the single-particle discrete Brownian motion system and introduced the concept of the equivalence of priors and constraints. In multi-particle systems, the equivalence of prior information and constraints extends beyond limitations on particle positions to include constraints on other physical quantities such as momentum and energy. In multi-particle systems, we continue to discuss maximum entropy distributions in closed systems and Self-Organization behavior in open systems through the lens of information, introducing concepts such as energy, Information Temperature, and Equilibrium Flow.

\subsection{Energy as Encoding and Information Temperature}

\subsubsection{Energy as Encoding}
In a multi-particle discrete Brownian motion system, to delve deeper into the interactions between particles and the system's dynamic behavior, we introduce the concept of Energy. Energy is not only a fundamental physical quantity in physical systems but also has an \textbf{informational essence}, directly determining the system's prior probability distribution and serving as a means of \textbf{information encoding}. The system's thermodynamic entropy \(S'\) and information entropy \(S\) are respectively defined as:

\begin{align}
    S' = -k_B \sum_i P(x_i) \ln P(x_i)   \\
    S = \sum_i P(x_i) \log_2 P(x_i)
\label{Thermodynamic Entropy and Information Entropy}
\end{align}
Here, \(P(x_i)\) is the probability of the microscopic state \(x_i\). From the perspective of Information Theory, the probability distribution \(P(x_i)\) determines the system's information entropy \(S\). Using a logarithmic base change, the relationship between thermodynamic entropy \(S'\) and information entropy \(S\) is given by:

\begin{equation}
    S' = k_B \ln 2 \cdot S
\end{equation}
To avoid ambiguity, we use expected information entropy instead of system information entropy for clarity, ignoring the coefficient of thermodynamic entropy in subsequent discussions. Information entropy represents the expected amount of information in the system, with expected information entropy \(S\) reflecting the degree of uncertainty in the system and the average amount of information required to describe its state. The higher the uncertainty of the system, the higher the information required to eliminate the uncertainty. Inspired by Landauer's principle, the coding length \(l_i\) and the probability of state occurrence \(P(x_i)\) are related by:

\begin{equation}
    l_i = -\ln P(x_i)
    \label{CodeLength}
\end{equation}
where \(P(x_i)\) follows a Boltzmann distribution: specifically, the energy \(E_i\) of each microscopic state \(x_i\) in the system is closely related to its probability \(P(x_i)\) and follows the Boltzmann distribution:

\begin{equation} 
    P(x_i) = \frac{1}{Z} \exp\left( -\frac{E_i}{k_B T} \right) 
    \label{Boltzmann Distribution} 
\end{equation}
Here, \(Z\) is the partition function, \(k_B\) is the Boltzmann constant, and \(T\) is the temperature. This means that the higher the energy of an event, the lower its probability, and the longer the coding length required. Substituting equation \eqref{Boltzmann Distribution} into the expression for coding length, we obtain the relationship between coding length and energy:

\begin{align}
    l_i &= -\ln \left( \frac{1}{Z} \exp\left( -\frac{E_i}{k_B T} \right) \right) \\
        &= \ln Z + \frac{E_i}{k_B T}
    \label{CodeLengthEnergy}
\end{align}

Thus, regard the temperature as a constant. The coding length \(l_i\) is proportional to energy \(E_i\). Higher energy corresponds to a longer coding length, indicating that energy influences the state probability and thus the coding length. Therefore, energy can be seen as a form of information encoding for the system. Therefore, \textbf{energy can be regarded as a measure of uncertainty reduction}, further emphasizing the \textbf{informational essence of energy}. The distribution of energy determines the observability of the system. From an informational perspective, energy not only describes the dynamical properties of particles but also serves as a means of encoding information. In the formula \eqref{CodeLengthEnergy}, the length of encoding is still related to the temperature. We think that temperature represents the frequency of sample information exchange between systems, so we propose \textbf{Information Temperature}.

\subsubsection{Information Temperature}
In this section, we introduce the concept of Information Temperature. Information Temperature represents the frequency and intensity of information exchange within a system. From the perspective of Information Theory, temperature can be interpreted as a measure of the frequency and intensity of information exchange within a system. In Thermodynamics, temperature characterizes the average kinetic energy of particles in a system and is closely related to the concepts of energy and thermodynamic entropy. Thermodynamic entropy is defined as:

\begin{equation}
    S' = k_B\ln{W}
\end{equation}
Here, \(W\) represents the number of microstates, and \(S'\) denotes thermodynamic entropy. In statistical mechanics, \(W\) is often defined as the total number of possible microstates of a system under given macroscopic conditions. For a system composed of many particles, each particle can occupy different positions and momentum states, creating different microstates. We analogize each state to partial information, with the probability of occurrence as \(p_i\), redefining thermodynamic entropy in terms of Information Theory \eqref{Thermodynamic Entropy and Information Entropy}. In Thermodynamics, temperature is defined as:

\begin{equation}
    \frac{1}{T}=\left(\frac{\partial S'}{\partial E}\right)_{V,N}
\end{equation}
Here, \(S'\) denotes thermodynamic entropy, \(E\) represents internal energy, and the derivative is taken under constant volume \(V\) and particle number \(N\). Given that \(S'\) is proportional to the dimensionless expected information entropy \(S\), Information Temperature is defined as:

\begin{equation}
    \frac{1}{T_{info}}=\frac{\partial(S)}{\partial\langle E\rangle}
\end{equation}
Here, \(\langle E \rangle\) denotes the system's average energy. $T_{info}$ represent the Information Temperature. The right side of the equation reflects the rate at which information entropy changes with average energy, quantifying the impact of energy changes on entropy. The expected information entropy of a system reflects its degree of uncertainty. Temperature, as a measure of information exchange, directly affects the system's information entropy. Higher temperatures imply more intense and frequent movement of microscopic particles, resulting in greater information exchange frequency and intensity, and consequently higher system information entropy. By substituting \(T_{info}\) into \eqref{Thermodynamic Entropy and Information Entropy}, the system's thermodynamic entropy becomes:

\begin{equation}
    S' = -\sum P(X(t))\cdot\ln P(X(t))= H=\frac{\langle E\rangle}{k_B\cdot T_{info}}+\ln Z\
    \label{Entropy_energy}
\end{equation}
By redefining temperature, we unify information entropy, energy, and temperature within a single framework, revealing their interconnections. Since \(T_{info}\) reflects the frequency and intensity of information exchange, it may vary dynamically with the system's information state. This implies that the prior distribution \(P(X(t))\) is also dynamic, reflecting the current information state of the system more accurately. In Bayesian updating, the posterior distribution depends on the prior distribution and the likelihood function. When the prior distribution incorporates Information Temperature, the posterior distribution is influenced by both microscopic information exchange and macroscopic information state, leading to an evolution of system entropy over time expressed as:

\begin{equation}
    \frac{dS'}{dt}=\frac{d}{dt}\left(\frac{\langle E(t)\rangle}{k_B\cdot T_{info}}+\ln Z\right)
\end{equation}
Thus, in multi-particle systems, the average energy \(\langle E(t) \rangle\) and Information Temperature \(T_{info}\) jointly determine the system's prior distribution. Higher Information Temperatures lead to more frequent information exchange among particles, potentially reducing the shielding effect of interactions and resulting in more disordered particle motion. The introduction of Information Temperature implies that information transmission within the system is influenced not only by particle interactions but also by the frequency and intensity of information exchange.

\subsubsection{Physical Meaning of Energy as Encoding}
We claim the view that the vacuum state, rather than being a simple void, is a highly complex and information-rich entity. It is precisely because of its infinite entropy that it becomes an unobservable background, which is in sharp contrast to the traditional perception that the vacuum is empty and thus unobservable.
In quantum field theory, the vacuum state is the lowest energy ground state. Although no real particles are present in the traditional sense, it is teeming with quantum fluctuations and the incessant creation and annihilation of virtual particle-antiparticle pairs. These quantum fluctuations give rise to a web of quantum entanglement.
In quantum mechanics, the entanglement entropy serves as a key metric for quantifying the degree of entanglement between subsystems within a quantum system. Consider a quantum system partitioned into regions A and B, with the overall state being a pure state $|\Psi\rangle$. The entanglement entropy of region A is defined as the von Neumann entropy of its corresponding subsystem:
\begin{equation}
S_A = -\text{Tr}(\rho_A \ln \rho_A)
\end{equation}
where \[ \rho_A = \text{Tr}_B |\Psi\rangle \langle \Psi| \] represents the normalized density matrix of region A. In the context of the vacuum state of a quantum field, it is far from being empty. For instance, consider a free real scalar field  in a d-dimensional space. When we divide this space into regions A and B and focus on the modes in the vicinity of the boundary of region A, we find that the entanglement entropy $S_A$ predominantly stems from high-frequency (short-wavelength) modes near the boundary. In the case of spatial dimensions $d \geq 3$, as the high-frequency cutoff $k_{max}$ is increased, the entanglement entropy $S_A$ diverges. This behavior is generally described by the Area Law:

\begin{equation}
S_A \propto A \cdot k_{\text{max}}^{d - 2}
\end{equation}
As $k_{max}$ approaches infinity, $S_A$ also tends to infinity. Given the established positive correlation between system entropy and expected information entropy, it follows that the expected information entropy of the system is likewise infinite. The infinite expected information entropy of the vacuum state implies that it is impossible to dispel its uncertainty with finite information. This is the fundamental reason why the vacuum remains an unobservable background. We are unable to obtain sufficient information to characterize its state. This conclusion aligns with the known physical properties of the vacuum: while the vacuum itself cannot be directly observed, its quantum fluctuations have a profound impact on physical processes. A prime example is the Casimir effect, which arises from vacuum fluctuations. When energy is introduced into the system, particles transition from virtual to real states and become observable. The injection of energy mitigates the uncertainty of the system, thereby transforming it from an unobservable background into an observable physical phenomenon.

\subsection{Open Systems and Self-Organization}

\subsubsection{Closed Systems}
To explore this principle from the perspective of Information Theory, we consider the rate of change of system entropy \( S \) over time. The total derivative of the expected entropy is given by:

\begin{equation}
\frac{dS}{dt} = \left( \frac{\partial S}{\partial \langle E \rangle} \right) \frac{d\langle E \rangle}{dt} + \left( \frac{\partial S}{\partial t} \right)
\label{EntropyTimeDerivative}
\end{equation}

In closed systems, the total energy of the system is conserved. The system's entropy does not decrease over time, and based on the positive correlation between system entropy and expected information entropy, the time rate of change of expected information entropy satisfies:

\begin{equation}
\frac{dS}{dt} \geq 0
\label{SecondLaw}
\end{equation}

\begin{equation}
\frac{d\langle E \rangle}{dt} = 0
\label{EnergyConservation}
\end{equation}
Therefore, in a closed system, the time rate of change of entropy is given by:

\begin{equation}
    \frac{dS}{dt} = \left( \frac{\partial S}{\partial t} \right) \geq 0
\end{equation}

\subsubsection{Information Temperature in Open Systems}
However, in non-equilibrium open systems, as the average energy of the system increases, the expected information entropy may decrease. This phenomenon is closely related to the concept of negentropy. The total entropy of the system does not violate the second law of Thermodynamics, as the system expels entropy to its environment, reducing its own expected information entropy, which can lead to Self-Organization and adaptive behavior. However, defining the specific form of such negentropy within the system, based on \textbf{Energy as Encoding} and \textbf{Information Temperature}, we propose the concepts of \textbf{Negative Information Temperature} and \textbf{Equilibrium Flow}. Negative Information Temperature is defined as the case where the Information Temperature takes on a negative value, under the following conditions:

\begin{enumerate}
    \item The system has an upper energy limit, allowing the Information Temperature \( T_{\text{info}} \) to assume negative values. \\ 
    \item The system's average energy increases over time, i.e., \(\mathrm{d}\langle E\rangle/\mathrm{dt} > 0\). \\ 
    \item Given the average energy, the explicit time dependence of the expected information entropy \( S \) can be ignored, i.e., \(\left( \partial S / \partial t \right)_{\langle E \rangle} = 0\). This implies that changes in entropy result from self-regulation and energy changes within the system, rather than direct time dependence. \\
\end{enumerate}
The constraint of an upper energy limit represents the maximum internal energy of the system, meaning the system cannot absorb external energy indefinitely. This restriction enables the system to self-regulate in response to external changes, preventing excessive energy accumulation or depletion that could lead to structural collapse, thereby maintaining stable structures. An increasing average energy over time indicates continuous energy acquisition from the environment, driving the system towards higher energy states. The constant energy input supports Self-Organization within the system, ensuring sufficient "drive" to maintain dynamic balance. Internally, the system must allocate its energy distribution freely as energy increases within a limited range, enhancing its efficiency in processing information. Given a specific average energy, the negligible explicit time dependence of expected information entropy implies that its change mainly relies on dynamic energy input and internal self-regulation, rather than direct time dependence. Time's influence on expected information entropy is a higher-order term and can be ignored.According to non-equilibrium Thermodynamics, the rate of change of expected information entropy can be decomposed into the entropy production rate \(\sigma\) and the entropy flux rate \(\Phi\):

\begin{equation}
    \frac{\mathrm{d}S}{\mathrm{d}t} = \sigma + \Phi
\end{equation}
Due to the second law of Thermodynamics, \(\sigma \geq 0\) represents the entropy production rate, indicating entropy generated by irreversible processes within the system; \(\Phi\) represents the entropy flux rate, indicating the exchange of entropy between the system and its environment. Information Temperature measures the frequency and intensity of information exchange within the system and is defined as the reciprocal of the partial derivative of expected information entropy \(S\) with respect to the system's average energy \(\langle E \rangle\). Negative Information Temperature occurs when \(\frac{\mathrm{d}S}{\mathrm{d}t} < 0\), implying \(\Phi < 0\), i.e., the system expels entropy to the environment, resulting in the presence of \textbf{negentropy}.

\subsubsection{Explaining Self-Organizing Behavior in Complex Systems}
Under the concept of Negative Information Temperature and the assumptions we have outlined, self-organizing phenomena can be interpreted as the presence of negentropy enabling the system to resist the trend of entropy increase, forming ordered structures and exhibiting \textbf{self-organizing behavior}. Open systems achieve the transition from disorder to order through energy input and entropy output. We use Energy as Encoding and Information Temperature to explain the self-organizing behavior of LLMs during training.
The training process of LLMs can be viewed as a standard form of this self-organizing behavior. Thanks to the development of big data and the liberation of computational power, LLMs have achieved unprecedented success by continuously expanding model size and datasets within excellent frameworks (e.g., transformers). Firstly, LLMs satisfy our three assumptions:
\begin{enumerate}
    \item The system has an upper energy limit. For LLMs, this corresponds to the maximum compressible information capacity of 2 Bits/Params\cite{allen2024physics}.
    \item The system's average energy increases over time, as the model interacts with increasing amounts of data, leading to a gradual energy increase within the system.
    \item Since the structure of LLMs remains nearly fixed once training begins, with only internal parameters changing, the explicit dependence of expected information entropy \(S\) on time remains constant at zero under given average energy conditions.
\end{enumerate}
Typical LLMs, such as GPT-3 and GPT-4, contain billions or even hundreds of billions of parameters. These parameters are randomly initialized at the beginning of training, with the initial distribution being far from the desired final distribution. As the dataset continues to be input, the model's internal energy increases. At the start of training, the model's internal energy is very low and far from the capability threshold, causing a rapid decrease in the loss value as it learns from the dataset. However, as energy approaches the threshold, external data continues to be input for training. The model must find more efficient ways to represent data, reducing the energy needed per unit of information and storing more information. At this stage, the model transitions from a low-energy state to a high-energy state, encoding information more effectively. This explains why \textbf{Energy as Encoding}. This self-regulatory behavior ensures that as energy increases, the model does not exceed its energy limit. During this phase, the loss value of LLMs decreases slowly. The gradual decrease in loss is accompanied by fluctuations, as certain data inputs cause temporary transitions to higher energy states, even if these may be suboptimal from a long-term perspective, explaining why the loss value fluctuates with continuous data input.

\subsection{Equilibrium Flow}
We name this framework, which explains self-organizing behavior based on \textbf{energy-as-coding} and \textbf{information temperature}, as \textbf{Equilibrium Flow}. Equilibrium Flow is defined as a state in non-equilibrium open systems where, under the constraint of an upper energy limit, continuous energy input (leading to increased average energy) and entropy flow output to the environment lead to reduced or stabilized internal entropy, forming a dynamic balance. In Complex System research, a Equilibrium Flow system represents a class of systems that achieve dynamic stability through energy input and negentropy output under non-equilibrium conditions. Such systems can spontaneously reach a state of self-organized criticality under specific conditions, exhibiting complex phenomena such as power-law distributions and high-frequency occurrence of high-energy events. To systematically describe and analyze the behavior of Equilibrium Flow systems, we propose three theorems:
\begin{enumerate}

    \item \textbf{Theorem 3: Equilibrium Flow Systems Follow the Fermi-Dirac Distribution:}

    In non-equilibrium open Complex Systems with an upper energy limit, if the system is under the condition of Negative Information Temperature \(T_{\text{info}} < 0\), the occurrence probability of events (such as particle occupation or specific state formation) \(P(E)\) follows the Fermi-Dirac distribution:  
    \begin{equation}
        P(E) = \frac{1}{\exp\left( \dfrac{\Phi_{\text{info}}(E)}{k_B T_{\text{info}}} \right) + 1}
    \end{equation}  
    where  \(\Phi_{\text{info}}(E)\) is the Information Alive Function of the system, represents the deviation of an event's energy from its occurrence threshold, determining the priority and likelihood of the event occurring.  \(k_B\) is the Boltzmann constant, and \(T_{\text{info}}\) is the Information Temperature. This distribution indicates a significantly increased occurrence probability of high-energy events under Negative Information Temperature conditions.

    \item 
    \textbf{Theorem 4: Self-Organized Criticality of Equilibrium Flow Systems:}

    Under the constraint of Negative Information Temperature \(T_{\text{info}} < 0\) and an upper energy limit, events within a Equilibrium Flow system (such as the occupation of high-energy states or the formation of specific structures) will spontaneously exhibit self-organized criticality. This criticality manifests as power-law distribution characteristics of event occurrences and cascading effects formed by local interactions, with the conditions for maintaining dynamic balance keeping the system near a critical state.
    
    \item \textbf{Theorem 5: Energy-Information Conversion Efficiency in Equilibrium Flow Systems:}

    Under Equilibrium Flow conditions, part of the energy input into the system is converted into information. The energy-to-information conversion efficiency \(\eta\) is given by the following relation:  
    \[
    \eta = \frac{\Delta I}{\Delta E} = \frac{1}{k_B |T_{\text{info}}|},
    \]  
    where \(\Delta I\) represents the increase in information, and \(\Delta E\) denotes the input energy. The conversion efficiency is inversely proportional to the absolute value of the Negative Information Temperature, i.e., the smaller the absolute value of the Negative Information Temperature, the higher the conversion efficiency.

\end{enumerate}

    



We now prove the above theorems. Formally, in a Equilibrium Flow system, let the system's energy input rate be \(\lambda_{\text{in}}\) and the dissipation rate be \(\lambda_{\text{diss}}\). When the system achieves dynamic balance, the following holds:

\begin{equation}
    \lambda_{\text{in}} = \lambda_{\text{diss}}.
\end{equation}
Under this condition, the system achieves dynamic stability through external energy input and negentropy output. The system evolves under conditions far from equilibrium, reaching a state of self-organized criticality. In Complex Systems, this behavior manifests as long-range correlations, scale invariance, and power-law distributions.

Theorem 3: \textbf{Equilibrium Flow Systems Follow the Fermi-Dirac Distribution}
In non-equilibrium open systems with an upper energy limit, if the system is under the condition of Negative Information Temperature \(T_{\text{info}} < 0\), the probability \(P(E)\) of event occurrence follows the Fermi-Dirac distribution:

\begin{equation}
    P(E) = \frac{1}{\exp\left( \dfrac{\Phi_{\text{info}}(E)}{k_B T_{\text{info}}} \right) + 1}
\end{equation}
Here, the \(\Phi_{\text{info}}(E)\) represent \textbf{Information Alive Function} :

\begin{equation}
    \Phi_{\text{info}}(E) = E - \mu
\end{equation}
\(\Phi_{\text{info}}(E)\) is the Information Alive Function of the system, representing the deviation between the event energy and the threshold for the event's occurrence. It determines the priority and probability of the event's occurrence. \(\mu\) is the chemical potential, representing the threshold for the event's occurrence. \(E\) denotes the "energy" of an event, characterizing the difficulty of its occurrence. The total information entropy of the system can be expressed as \eqref{Entropy_energy}. We introduce the constraint condition \textbf{normalization condition}:

\begin{equation}
   \sum_i P(E_i) = N
\end{equation}
where \(N\) denotes the total number of events or particles in the system. \textbf{Average energy constraint}:

\begin{equation}
   \sum_i P(E_i) E_i = \langle E \rangle
\end{equation}
Here, \(\langle E \rangle\) is the system's average energy. We construct the Lagrange function:

\begin{equation}
   \mathcal{L} = -k_B \sum_i \left[ P(E_i) \ln P(E_i) + \left( 1 - P(E_i) \right) \ln \left( 1 - P(E_i) \right) \right] - \alpha \left( \sum_i P(E_i) - N \right) - \beta \left( \sum_i P(E_i) E_i - \langle E \rangle \right)
\end{equation}
where \(\alpha\) and \(\beta\) are Lagrange multipliers. Taking the derivative with respect to \(P(E_i)\) and setting it to zero yields:

\begin{equation}
   \frac{\partial \mathcal{L}}{\partial P(E_i)} = -k_B \left[ \ln P(E_i) - \ln \left( 1 - P(E_i) \right) \right] - \alpha - \beta E_i = 0
\end{equation}
Rearranging gives:

\begin{equation}
   \ln \frac{P(E_i)}{1 - P(E_i)} = \frac{\Phi_{\text{info}}(E_i)}{k_B T_{\text{info}}}
\end{equation}
We use the following relations:

\begin{align}
   T_{\text{info}} &= -\frac{1}{\beta} \\
   \mu &= -\frac{\alpha}{\beta} k_B T_{\text{info}}
\end{align}
Thus, the probability distribution of events is given by:

\begin{equation}
    P(E_i) = \frac{1}{\exp\left( \dfrac{\Phi_{\text{info}}(E_i)}{k_B T_{\text{info}}} \right) + 1}
\end{equation}
This indicates that the event occurrence probability directly depends on the Information Alive Function \(\Phi_{\text{info}}(E_i)\) and follows the Fermi-Dirac distribution form. Since \(T_{\text{info}} = -\dfrac{1}{\beta} < 0\) and \(\mu = -\dfrac{\alpha}{\beta} k_B T_{\text{info}}\),  \(T_{\text{info}}\) are negative, implying an increased probability of high-energy events (\(E_i > \mu\)).

Theorem 4: \textbf{Self-Organized Criticality of Equilibrium Flow Systems} states that under the constraint of Negative Information Temperature \(T_{\text{info}} < 0\) and an upper energy limit, events within a Equilibrium Flow system (such as occupation of high-energy states or structure formation) spontaneously exhibit self-organized criticality. Consider a system consisting of \(N\) subunits, each with a maximum energy constraint \(E_{\text{max}}\). The state distribution of the system is influenced by the Negative Information Temperature \(T_{\text{info}} < 0\), leading to a higher probability of occupying high-energy states. Assuming an energy transfer mechanism within the system, when a subunit reaches a threshold \(E_c\), it triggers energy release. The system maintains balance between external energy input and internal dissipation, satisfying:

\begin{equation}
       \frac{dE_{\text{in}}}{dt} = \lambda_{\text{in}}, \quad \frac{dE_{\text{diss}}}{dt} = \lambda_{\text{diss}}
\end{equation}
under balance conditions, \(\lambda_{\text{in}} = \lambda_{\text{diss}}\). This ensures that the system does not accumulate energy indefinitely but fluctuates around a stable state. Through local interactions (e.g., energy transfer), the system forms complex cascading effects. Small local disturbances may trigger large-scale energy releases, with the event size \(s\) following a power-law distribution:

\begin{equation}
    P(s) \propto s^{-\tau}
\end{equation}
Thus, under the constraints of Negative Information Temperature and an upper energy limit, a system achieves self-organized criticality through the balance of energy input and dissipation, exhibiting self-organizing behavior. This scale invariance and long-range correlation are hallmarks of self-organized criticality.

Theorem 5: \textbf{Energy-Information Conversion Efficiency in Equilibrium Flow Systems} states that under Equilibrium Flow conditions, the energy-to-information conversion efficiency is inversely proportional to the absolute value of the Negative Information Temperature. The system converts part of the input energy into information, with the conversion efficiency \(\eta\) given by:

\begin{equation}
    \eta = \frac{\Delta I}{\Delta E} = \frac{1}{k_B |T_{\text{info}}|}
\end{equation}
where \(\Delta I\) denotes the increase in information, and \(\Delta E\) denotes the input energy. According to the relationship between entropy and information\eqref{Entropy_energy}, the total information \(I\) of the system and the information entropy \(S\) are related by:

\begin{equation}
   I = -\frac{S}{k_B}
\end{equation}
Thus, the change in information is inversely proportional to the change in expected information entropy:

\begin{equation}
    \Delta I = -\frac{\Delta S}{k_B}
\end{equation}
According to the definition of Information Temperature, we have:

\begin{equation}
   \frac{1}{T_{\text{info}}} = \frac{\partial S}{\partial \langle E \rangle}
\end{equation}
When the energy change is small, we can approximate:

\begin{equation}
   \Delta S \approx \frac{1}{T_{\text{info}}} \Delta E
\end{equation}
Substituting the above relationships into the expression for energy-information conversion efficiency, since \(T_{\text{info}} < 0\), the conversion efficiency is positive:

\begin{equation}
   \eta = \frac{\Delta I}{\Delta E} = -\frac{\Delta S}{k_B \Delta E} = -\frac{1}{k_B T_{\text{info}}} = \frac{1}{k_B |T_{\text{info}}|}
\end{equation}
Thus, the energy-to-information conversion efficiency is inversely proportional to the absolute value of the Negative Information Temperature; the smaller the absolute value, the higher the efficiency.

Thus, in Equilibrium Flow systems, the probability of events follows the Fermi-Dirac distribution form, indicating a higher probability of high-energy events under Negative Information Temperature, reflecting the self-organizing nature of the system. By defining and deriving the characteristics of Equilibrium Flow systems, we have demonstrated how Negative Information Temperature and energy limits drive Complex Systems to spontaneously reach self-organized critical states, improving information processing efficiency. Furthermore, the event probability distribution exhibits Fermi-Dirac distribution characteristics, providing a theoretical basis for understanding the statistical behavior of Complex Systems.

\section{Conclusion}
This article revisits Brownian motion from the perspective of Information Theory, introducing the single-particle discrete Brownian motion model (SPBM) to demonstrate the equivalence between prior information and constraint conditions, and revealing the interplay between local randomness and global probability distributions. Extending to multi-particle systems, it presents "Energy as Encoding" and "Information Temperature" to explain how energy distribution shapes information structures, and proposes the concept of "Equilibrium Flow" to describe self-organizing behaviors under energy constraints and Negative Information Temperature, applying these ideas to the training of large language models (LLMs). While the work offers a unified theoretical framework connecting Information Theory, Thermodynamics, and Complex Systems, it primarily provides theoretical insights without extensive empirical validation or detailed mathematical proofs, which may limit its immediate applicability. Future research could focus on rigorous mathematical formalization, empirical testing of the proposed concepts, and exploring practical applications in complex systems and artificial intelligence.

\bibliographystyle{unsrt}  
\bibliography{references}  

\end{document}